\documentclass[12pt]{iopart}
\usepackage{cite}
\usepackage{iopams}
\usepackage{graphicx}

\begin{document}

\title{Scaling properties of signals as origin of $1/f$ noise}

\author{J Ruseckas and B Kaulakys}

\address{Institute of Theoretical Physics and Astronomy, Vilnius University,
A. Go\v{s}tauto 12, LT-01108 Vilnius, Lithuania}
\ead{julius.ruseckas@tfai.vu.lt}

\begin{abstract}
There are several mathematical models yielding $1/f$ noise. For example, $1/f$
spectrum can be obtained from stochastic sequence of pulses having power-law
distribution of pulse durations or from nonlinear stochastic differential
equations. We show that a couple of seemingly different models exhibiting $1/f$
spectrum are due to the similar scaling properties of the signals. In addition,
we demonstrate a connection between signals with the power-law behavior of the
power spectral density generated by the nonlinear stochastic differential
equations and modeled by a sequence of random different pulses. An approximation
of solutions of the nonlinear stochastic differential equations by the sequence
of pulses correctly reproduces the power-law parts of the probability density
function and of the power spectral density. This connection provides further
insights into the origin of $1/f$ noise.

\noindent{\it Keywords\/}: stochastic processes (theory), current fluctuations,
stationary states
\end{abstract}

\maketitle

\section{Introduction}

Signals having the power spectral density (PSD) at low frequencies
$f$ of the form $S(f)\sim1/f^{\beta}$ with $\beta$ close to $1$
are commonly referred to as ``$1/f$ noise'', ``$1/f$ fluctuations'',
or ``flicker noise.'' Power-law distributions of spectra of signals
with $0.5<\beta<1.5$, as well as scaling behavior are
ubiquitous in physics and in many other fields \cite{Scholarpedia2007,Weissman1988,Barabasi1999,Gisiger2001,Wong2003,Wagenmakers2004,Newman2005,Szabo2007,Castellano2009,Eliazar2009,Eliazar2010,Perc2010,Orden2010,Kendal2011,Torabi2011,Diniz2011}.
Despite the numerous models and theories proposed since its discovery
more than 80 years ago \cite{Johnson1925,Schottky1926}, the subject
of $1/f$ noise remains still open for new discoveries. Most models
and theories of $1/f$ noise are not universal because of the assumptions
specific to the problem under consideration. A short categorization
of the theories and models of $1/f$ noise is presented in the introduction
of the paper \cite{Kaulakys2009}. 
See also recent experimental investigations and physical models of $1/f$ noise in condensed matter \cite{Liu2009,Dmitruk2011,Liu2013,Han2013,Kuhlmann2013} and a review by Balandin \cite{Balandin2013}. 

Often $1/f$ noise is modeled as the superposition of Lorentzian spectra
with a wide range distribution of relaxation times \cite{McWhorter1957}. An
influential class of the models of $1/f$ noise involves self-organized
criticality (SOC). In 1987 Bak \textit{et al} \cite{Bak1987} introduced
the notion of SOC with the motivation to explain the universality
of $1/f$ noise. Although paper \cite{Bak1987} is the most cited
paper in the field of $1/f$ noise problems, it was shown later on
\cite{Jensen1989,Kertesz1990} that the mechanism proposed in \cite{Bak1987}
results in $1/f^{\beta}$ fluctuations with $1.5<\beta\leq2$ and does
not explain the omnipresence of $1/f$ noise. The $1/f$ noise in
the fluctuations of mass was first seen in a sandpile model with threshold
dissipation by Ali \cite{Ali1995}. Maslov \textit{et al} \cite{Maslov1999}
studied a one-dimensional directed model of sandpiles and showed that
the exponent $\beta$ is exactly $1$ for noise in the spectrum of
fluctuations of mass. An example of $1/f$ noise in the classical
sandpile model has been provided in \cite{Baiesi2006}. 

Another models of $1/f$ noise involve a class of maps generating intermittent
signals. It is possible to generate power-laws and $1/f$-noise from simple
iterative maps by fine-tuning the parameters of the system at the edge of chaos
\cite{Procaccia1983,Schuster1988} where the sensitivity to initial conditions of
the logistic map is a lot milder than in the chaotic regime \cite{Costa1997}.
Manneville \cite{Manneville1980a} showed that iterative function can produce
interesting behavior, power-laws and $1/f$ PSD. In paper
\cite{RuseckasChaos2013} a mechanism of intermittency exhibiting $1/f$ noise
which occurs in nonlinear dynamical systems with invariant subspace and having
the transverse Lyapunov exponent equal to zero was considered. Intermittency as
a mechanism of $1/f$ noise continues to attract attention
\cite{Laurson2006,Pando2007,Shinkai2012}. 

In many cases the physical processes can be represented by a sequence
of random pulses. The mathematical way of generating power-law noise
from a sequence of pulse has been discussed by Halford \cite{Halford1968}.
The spectrum of the signal consisting of the pulse sequences which
belong to the class of Markov process, was investigated in \cite{Heiden1969,Shick1974}.
In those models the power-law PSD appears due to power-law distribution
of pulse durations. Main objection to this approach is an apparent
lack of physical mechanism generating durations over several orders
of magnitude \cite{Press1978}. On the other hand, we can point out
that the diffusion process gives rise to broad distributions of lifetimes.
It is known that for the unbiased random walk, the distribution of
the first return times has a power-law form with the exponent $-3/2$
\cite{Redner2001}. Another possible mechanism leading to broad distribution
of lifetimes is owing to the formation of avalanches. In many situations (for example
in Barkhausen noise \cite{Kuntz2000}, fluid invasion into disordered
media \cite{Rost2007}, dislocation avalanches in microcrystals
\cite{Papanikolaou2012}) the volume of avalanches has power-law distribution
and, in addition, the volume scales as a power-law function of the
duration of the avalanche. Similar properties have also avalanches
in the models of SOC \cite{Freeman2000,Laurson2005,Bartolozzi2006,Yadav2012,LeBlanc2013}. 

One more way of obtaining $1/f$ noise from a signal consisting of
pulses has been presented in \cite{Kaulakys1998,Kaulakys1999,Kaulakys2000-2,Kaulakys2005}.
It has been shown that the intrinsic origin of $1/f$ noise may be
a Brownian motion of the interevent time of the signal pulses, similar
to the Brownian fluctuations of the signal amplitude, resulting in
$1/f^{2}$ noise. Recently, the nonlinear stochastic differential
equations (SDEs) generating signals with $1/f$ noise were obtained
in \cite{Kaulakys2004,Kaulakys2006} (see also papers \cite{Kaulakys2009,Ruseckas2010,ruseckas-2011}),
starting from the point process model of $1/f$ noise. 
Analysis of the long-range correlated bursting signals is presented in \cite{Davidsen2009,Grigolini2010,Moloney2011,Martin2011,Gontis2012ACS,Eliazar2013,Krisponeit2013}, as well. 

The purpose of this paper is to show the connection between the nonlinear SDEs
generating signals with $1/f$ noise and signals consisting of random pulses
with the power-law distribution of pulse durations. As we will demonstrate, in
both of these models $1/f$ spectrum appears due to the scaling properties of the
signal. In addition, the signal generated by SDE can be approximated by rectangular
pulses yielding the same distribution of signal intensity and the same power-law
exponent in the PSD. Although the models generating $1/f$ noise that we consider
in this paper (nonlinear SDEs and random pulses with the power-law distribution
of pulse durations) are not unique, the method of their derivation from the scaling
properties of the signal was not investigated before. We obtain nonlinear SDEs
generating signals with $1/f$ noise starting {\itshape not} from the point process model,
as has been done in \cite{Kaulakys2004,Kaulakys2006}, but {\itshape from the scaling properties of
the signal required to get $1/f$ noise}. This approach allows us to reveal new
connections between those seemingly different models.

The paper is organized as follows: In \sref{sec:SDE} we consider
nonlinear SDEs generating signals with $1/f^{\beta}$ PSD and show
that such SDEs can be obtained by requiring a proper scaling. In
\sref{sec:pulses} we analyze signals consisting of random pulses with the 
power-law distribution of pulse durations and power-law dependence
of pulse height on the pulse duration. We show that such pulses have
the same scaling properties as the signal generated by SDEs in \sref{sec:SDE}.
In \sref{sec:sde-pulses} we produce the connection between the nonlinear
SDEs modeling and the rectangular pulses series more explicitly. \Sref{sec:concl}
summarizes our findings.

\section{Nonlinear SDE generating signals with $1/f^{\beta}$ noise from scaling}

\label{sec:SDE}Nonlinear SDEs generating signals with $1/f^{\beta}$ PSD are
derived in papers \cite{Kaulakys2004,Kaulakys2006}. In this Section we show these SDEs
can be obtained only from the scaling properties required by $1/f^{\beta}$ PSD.
This new technique reveals more directly the origin of $1/f^{\beta}$ PSD compared to
the derivation starting from the point process model, as has been done in
\cite{Kaulakys2004,Kaulakys2006}.

Pure $1/f^{\beta}$ PSD is physically impossible because the total
power would be infinite. Therefore, we will consider signals with PSD
having $1/f^{\beta}$ behavior only in some wide intermediate region
of frequencies, $f_{\mathrm{min}}\ll f\ll f_{\mathrm{max}}$, whereas
for small frequencies $f\ll f_{\mathrm{min}}$ PSD is bounded. We
can obtain nonlinear SDE generating signals exhibiting $1/f$ noise
using the following considerations. Wiener-Khintchine theorem relates
PSD $S(f)$ to the autocorrelation function $C(t)$: 
\begin{equation}
C(t)=\int_{0}^{+\infty}S(f)\cos(2\pi ft)\rmd f\,.\label{eq:W-K}
\end{equation}
If $S(f)\sim f^{-\beta}$ in a wide region of frequencies, then for
the frequencies in this region the PSD has a scaling property 
\begin{equation}
S(af)\sim a^{-\beta}S(f)
\end{equation}
when the influence of the limiting frequencies $f_{\mathrm{min}}$
an $f_{\mathrm{max}}$ is neglected. From the Wiener-Khintchine theorem
\eref{eq:W-K} it follows that the autocorrelation function has the
scaling property 
\begin{equation}
C(at)\sim a^{\beta-1}C(t)\label{eq:auto-scaling}
\end{equation}
in the time range $1/f_{\mathrm{max}}\ll t\ll1/f_{\mathrm{min}}$.
The autocorrelation function can be written as
\cite{Ruseckas2010,Risken1996,Gardiner2004}
\begin{equation}
C(t)=\int\rmd x\int\rmd x'\, xx'P_{0}(x)P(x',t|x,0)\,,\label{eq:auto}
\end{equation}
where $P_{0}(x)$ is the steady state probability density function
(PDF) and $P(x',t|x,0)$ is the transition probability (the conditional
probability that at time $t$ the signal has value $x'$ with the
condition that at time $t=0$ the signal had the value $x$). The
transition probability can be obtained from the solution of the Fokker-Planck
equation with the initial condition $P(x',t|x,0)=\delta(x'-x)$.
One of the ways to obtain the required property \eref{eq:auto-scaling} is
for the steady state PDF to have the power-law form
\begin{equation}
P_{0}(x)\sim x^{-\lambda}\label{eq:prob-steady}
\end{equation}
and for the transition probability to have the scaling property 
\begin{equation}
aP(ax',t|ax,0) = P(x',a^{\mu}t|x,0)\,,\label{eq:prob-scaling}
\end{equation}
where $\mu$ is the scaling exponent, the meaning of which will be revealed below,
equation \eref{eq:mu}.  
Indeed, from equations \eref{eq:auto}--\eref{eq:prob-scaling} and a change of
variables it follows
\begin{eqnarray}
C(at) & = \int\rmd x\int\rmd x'\, xx'P_{0}(x)P(x',at|x,0) \\
& \sim \int\rmd x\int\rmd x'\, x^{1-\lambda}x'a^{\frac{1}{\mu}}
P(a^{\frac{1}{\mu}}x',t|a^{\frac{1}{\mu}}x,0) \\
 & \sim a^{\frac{\lambda-3}{\mu}}\int\rmd u\int\rmd u'\, uu'P_{0}(u)P(u',t|u,0)\,.
\end{eqnarray}

Thus, the autocorrelation function has the required property
\eref{eq:auto-scaling} with $\beta$ given by equation
\begin{equation}
\beta = 1 + (\lambda-3)/\mu\,.\label{eq:beta-sde-1}
\end{equation}
Note, that according to equation \eref{eq:prob-scaling} the change of the
magnitude of the stochastic variable $x\rightarrow ax$ is equivalent to the change of time
scale $t\rightarrow a^\mu t$.

In order to avoid the divergence of steady state PDF \eref{eq:prob-steady}
the diffusion of stochastic variable $x$ should be restricted at
least from the side of small values and, therefore, \eref{eq:prob-steady}
holds only in some region of the variable $x$, $x_{\mathrm{min}}\ll x\ll x_{\mathrm{max}}$.
When the diffusion of stochastic variable $x$ is restricted, equation \eref{eq:prob-scaling}
also cannot be exact. However, if the influence of the limiting values
$x_{\mathrm{min}}$ and $x_{\mathrm{max}}$ can be neglected for time
$t$ in some region $t_{\mathrm{min}}\ll t\ll t_{\mathrm{max}}$,
we can expect that the scaling \eref{eq:auto-scaling} approximately holds
in this time region.

To get the required scaling \eref{eq:prob-scaling} of the transition
probability, the SDE should contain only powers of the stochastic variable $x$.
This will be the case if the coefficient in the noise term is the power-law depending, i.e., proportional to
$x^{\eta}$. The drift term then is fixed by the requirement
\eref{eq:prob-steady} for the steady state PDF. Thus we consider SDE \cite{Kaulakys2009}
\begin{equation}
\rmd x=\sigma^{2}\left(\eta-\frac{1}{2}\lambda\right)x^{2\eta-1}\rmd t
+\sigma x^{\eta}\rmd W_t\,.\label{eq:sde}
\end{equation}
Here $W_t$ is a standard Wiener process (the Brownian motion) and $\sigma$ is
the white noise intensity. Note that SDE \eref{eq:sde} is the same as in papers 
\cite{Kaulakys2009,Kaulakys2006}, only here we obtained it from the
consideration of the scaling properties, not starting from the point process model.
Changing the variable $x$ in \eref{eq:sde} to the
scaled variable $x_{\mathrm{s}}=ax$ or introducing the scaled time
$t_{\mathrm{s}}=a^{2(\eta-1)}t$ and using the property of the Wiener process
$\rmd W_{t_{\mathrm{s}}}=a^{\eta-1}\rmd W_t$ one gets the same resulting equation.
Thus, change of the scale of the variable $x$ and change of time
scale are equivalent, as in equation \eref{eq:prob-scaling},
and the exponent $\mu$ is 
\begin{equation}
\mu=2(\eta-1)\,.\label{eq:mu}
\end{equation}
From equation \eref{eq:beta-sde-1} it follows that the power-law
exponent in the PSD of the signal generated by SDE \eref{eq:sde} is
\begin{equation}
\beta=1+\frac{\lambda-3}{2(\eta-1)}\,.\label{eq:beta-sde}
\end{equation}

In order to obtain a stationary process and avoid the divergence of
steady state PDF the diffusion of stochastic variable $x$ should
be restricted or equation \eref{eq:sde} should be modified. The
simplest choice of the restriction is the reflective boundary conditions
at $x=x_{\mathrm{min}}$ and $x=x_{\mathrm{max}}$. Another choice would be
modification of equation \eref{eq:sde} to get rapidly decreasing steady state PDF
when the stochastic variable $x$ acquires values outside of the interval
$[x_{\mathrm{min}}, x_{\mathrm{max}}]$. For example, the steady state PDF
\begin{equation}
P_{0}(x)\sim\frac{1}{x^{\lambda}}\exp\left\{
-\left(\frac{x_{\mathrm{min}}}{x}\right)^{m}
-\left(\frac{x}{x_{\mathrm{max}}}\right)^{m}\right\} 
\end{equation}
with $m>0$ has a power-law form when $x_{\mathrm{min}}\ll x\ll x_{\mathrm{max}}$
and exponential cut-offs when $x$ is outside of the interval $[x_{\mathrm{min}}, x_{\mathrm{max}}]$.
Such exponentially restricted diffusion is generated by the SDE 
\begin{equation}
\rmd x=\sigma^{2}\left[\eta-\frac{1}{2}\lambda
+\frac{m}{2}\left(\frac{x_{\mathrm{min}}^{m}}{x^{m}}
-\frac{x^{m}}{x_{\mathrm{max}}^{m}}\right)\right]x^{2\eta-1}\rmd t
+\sigma x^{\eta}\rmd W_t\label{eq:sde-restricted}
\end{equation}
obtained from equation \eref{eq:sde} by introducing additional terms in the drift.

The presence of the restrictions at $x=x_{\mathrm{min}}$ and $x=x_{\mathrm{max}}$
makes the scaling \eref{eq:prob-scaling} not exact and this limits
the power-law part of the PSD to a finite range of frequencies
$f_{\mathrm{min}}\ll f\ll f_{\mathrm{max}}$.
Let us estimate the limiting frequencies. Taking into account the
limiting values $x_{\mathrm{min}}$ and $x_{\mathrm{max}}$, equation \eref{eq:prob-scaling}
for the transition probability corresponding to SDE \eref{eq:sde}
becomes 
\begin{equation}
aP(ax',t|ax,0;ax_{\mathrm{min}},ax_{\mathrm{max}})=
P(x',a^{\mu}t|x,0;x_{\mathrm{min}},x_{\mathrm{max}})\,.\label{eq:prob-scaling-bound}
\end{equation}
Here $x_{\mathrm{min}}$, $x_{\mathrm{max}}$ are the parameters of the transition
probability. The steady state distribution
$P_{0}(x;x_{\mathrm{min}},x_{\mathrm{max}})$ has the scaling property 
\begin{equation}
aP_{0}(ax;ax_{\mathrm{min}},ax_{\mathrm{max}})=
P_{0}(x;x_{\mathrm{min}},x_{\mathrm{max}})\,.\label{eq:prob-steady-bound}
\end{equation}
Inserting equations \eref{eq:prob-scaling-bound} and \eref{eq:prob-steady-bound}
into equation \eref{eq:auto} we obtain 
\begin{equation}
C(t;ax_{\mathrm{min}},ax_{\mathrm{max}})=
a^{2}C(a^{\mu}t,x_{\mathrm{min}},x_{\mathrm{max}})\,.
\end{equation}
This equation means that time $t$ in the autocorrelation function
should enter only in combinations with the limiting values,
$x_{\mathrm{min}}t^{1/\mu}$
and $x_{\mathrm{max}}t^{1/\mu}$. We can expect that
the influence of the limiting values can be neglected and the scaling \eref{eq:prob-scaling}
holds when the first combination is small and the second large, that
is when time $t$ is in the interval
$\sigma^{-2}x_{\mathrm{max}}^{-\mu}\ll t\ll\sigma^{-2}x_{\mathrm{min}}^{-\mu}$.
Then, using equation \eref{eq:W-K} the frequency range where the PSD
has $1/f^{\beta}$ behavior can be estimated as 
\begin{equation}
\sigma^{2}x_{\mathrm{min}}^{\mu}\ll2\pi f\ll
\sigma^{2}x_{\mathrm{max}}^{\mu}\,.\label{eq:freq-range}
\end{equation}
However, numerical solutions of proposed nonlinear SDEs show that
this estimation is too broad, i.e., the numerically obtained frequency region
with the power-law behavior of PSD is narrower than according to equation \eref{eq:freq-range}.
Note, that for $\mu=0$, i.e., $\eta=1$ the width of the frequency
region \eref{eq:freq-range}
is zero, and we do not have $1/f^{\beta}$ power spectral density.

\begin{figure}
\includegraphics[width=0.33\textwidth]{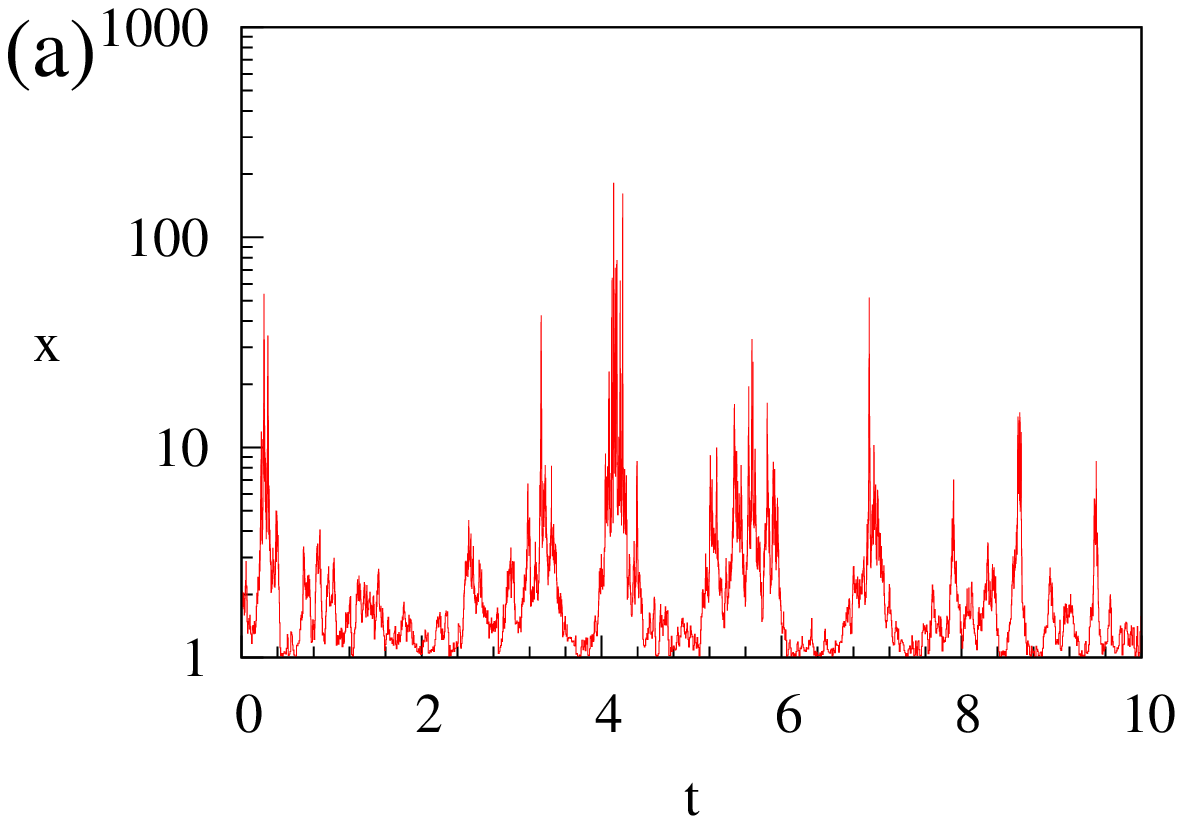}\includegraphics[width=0.33\textwidth]{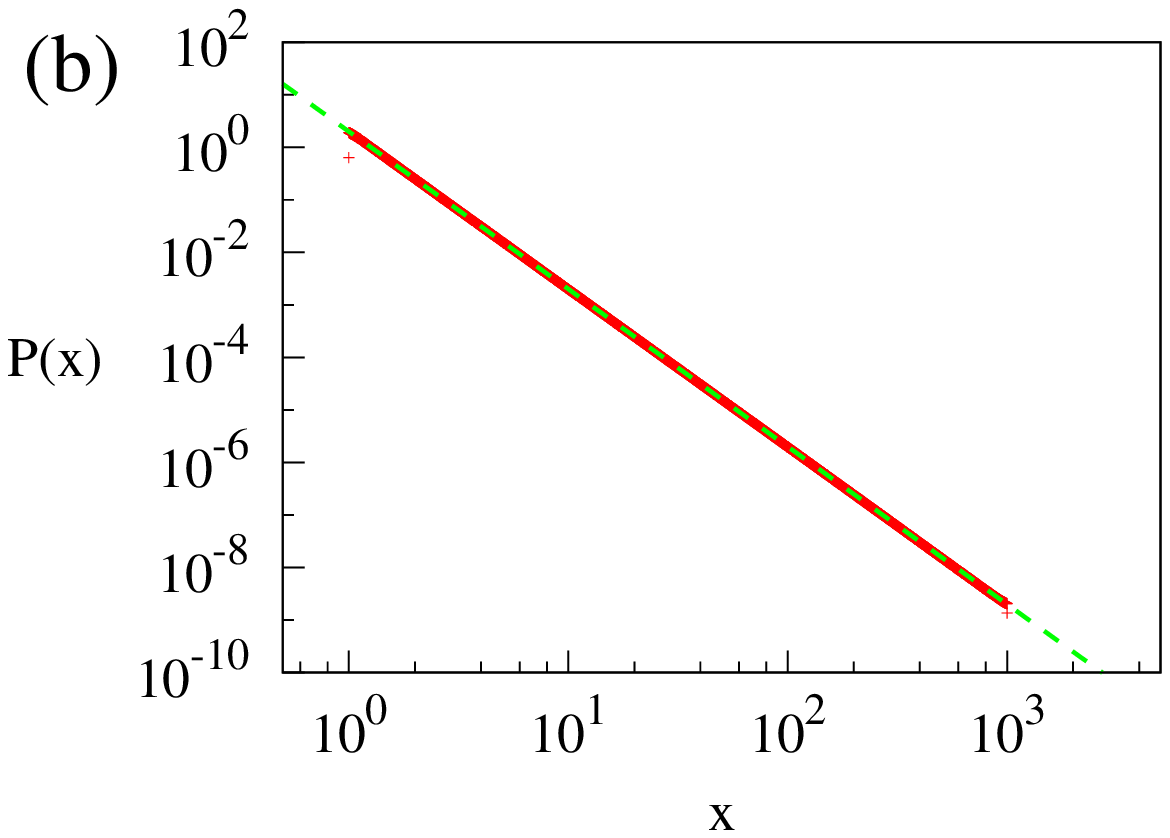}\includegraphics[width=0.33\textwidth]{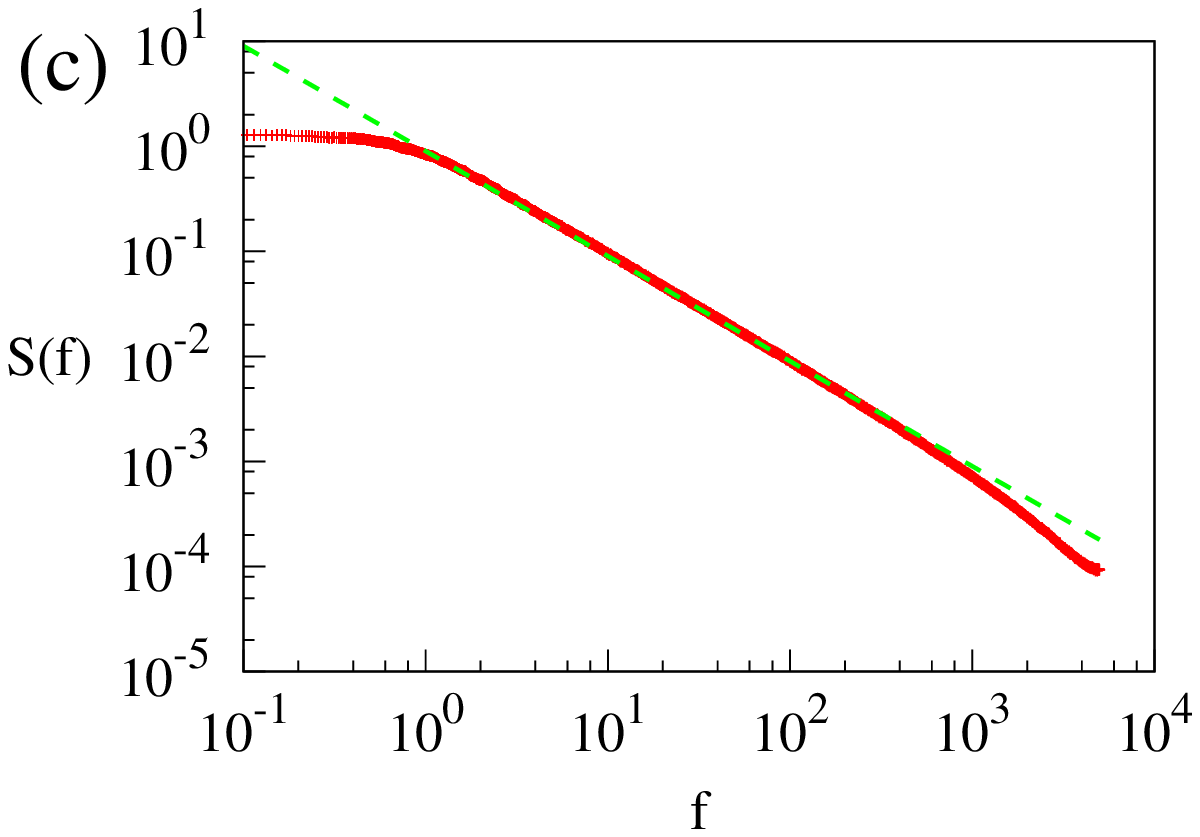}
\caption{(Color online) (a) Typical signal generated by equation \eref{eq:sde}
with reflective boundaries at $x_{\mathrm{min}}$ and $x_{\mathrm{max}}$. 
(b) The PDF of the signal intensity. The dashed (green) line shows
the power-law with the exponent $-3$. (c) The PSD of such a signal.
The dashed (green) line shows the slope $f^{-1}$. Used parameters
are $\eta=2$, $\lambda=3$, $x_{\mathrm{min}}=1$, $x_{\mathrm{max}}=1000$, and 
$\sigma=1$.}
\label{fig:sde}
\end{figure}

Comparison of the numerically obtained steady state PDF and the PSD with
analytical expressions for SDE \eref{eq:sde} with $\eta=\mu=2$ and $\lambda=3$
is presented in figure~\ref{fig:sde}. For the numerical solution we use the
Euler-Maruyama approximation, transforming the differential equations to
difference equations. We can use constant time step, however at large values of
$x$ the coefficients in the equations become large and thus require a very small
time step. More effective method of solution is to use a variable time step,
decreasing with the increase of $x$. As in \cite{Kaulakys2004,Kaulakys2006} we
choose the time step in such a way that the coefficient before noise becomes
proportional to the first power of $x$. Very similar numerical results one gets
also by using the Milstein approximation \cite{Kaulakys2009}. We see good
agreement of the numerical results with the analytical expressions. A numerical
solution of the equations confirms the presence of the frequency region for
which the power spectral density has $1/f^{\beta}$ dependence. The $1/f$
interval in the PSD in figure~\ref{fig:sde} is approximately between
$f_{\mathrm{min}}\approx10^{0}$ and $f_{\mathrm{max}}\approx10^{3}$ and is much
narrower than the width of the region $1\ll f\ll10^{6}$ predicted by equation 
\eref{eq:freq-range}. The width of this region can be increased by increasing
the ratio between the minimum and the maximum values of the stochastic variable
$x$.

As we see in figure~\ref{fig:sde}a, the numerical calculations exhibit a structure
of the signal consisting of peaks or bursts. Analysis \cite{Kaulakys2009}
reveals that the sizes of the bursts are approximately proportional to the
squared durations of the bursts with the power-law distributions of the bursts
durations and interburst time. The exponent of the PDF of the interburst time
approximately equal to $-3/2$ has been obtained numerically \cite{Kaulakys2009}
and analytically \cite{Gontis2012ACS}. 

\section{Stochastic pulse sequences}

\label{sec:pulses}
\begin{figure}
\includegraphics[width=0.4\textwidth]{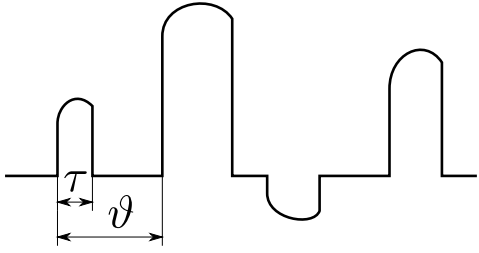}
\caption{Sequence of random pulses.}
\label{fig:pulses}
\end{figure}

In this Section we consider pulse sequences with
independent pulses. The shapes of the pulses are characterized by an arbitrary large
set of parameters $\xi$, whereas the occurrence times of the pulses are described by a set of
time moments $\{t_k\}$. The general form of the signal can be written
as
\begin{equation}
I(t)=\sum_{k}A(t-t_{k},\xi_{k})\,,\label{eq:signal}
\end{equation}
where functions $A(t,\xi)$ determine the shape of individual pulses. The pulse
duration $\tau$ is included in the set of parameters $\xi$ or, more
generally, is a function of the parameters, $\tau(\xi)$. Inter-pulse duration is
$\vartheta_{k}=t_{k+1}-t_{k}$. Such a pulse
sequence is schematically shown in figure~\ref{fig:pulses}. We assume
that: (i) the pulse sequences are stationary and ergodic; (ii) parameters
$\xi$ of different pulses are independent; (iii) all pulses 
are described by the same function $A(t,\xi)$; (iv) the pulse parameters
$\xi$ have the distribution $P(\xi)$. 

The easies way to calculate
PSD of a pulse sequence is to start from the definition of the power
spectral density directly \cite{Heiden1969}. However, in order to
demonstrate connection with the previous Section we will consider
the expression for the autocorrelation function of the signal.
The autocorrelation function is given by the equation
\begin{equation}
C(t)=\lim_{T\rightarrow\infty}\left\langle \frac{1}{T-t}\int_{0}^{T-t}I(t')I(t'+t)
\rmd t'\right\rangle \,,\label{eq:autocorr-pulses}
\end{equation}
where $T$ the observation time interval and the brackets $\langle\cdot\rangle$ denote
averaging over realizations of the pulse sequence. Using the signal
\eref{eq:signal} we can write
\begin{equation}
C(t)=\lim_{T\rightarrow\infty}\left\langle \frac{1}{T}\sum_{k,k'=1}^{N}
\int_{0}^{\infty}A(t',\xi_{k})A(t'+t+t_{k}-t_{k'},\xi_{k'})\rmd t'\right\rangle\,,
\end{equation}
where $N$ is the number of pulses during the observation time interval $T$. 

The autocorrelation function can be decomposed into two parts, the first part containing the
autocorrelation of each pulse with itself and the second part containing all cross terms: 
\begin{equation}
C(t)=\lim_{T\rightarrow\infty}\left\langle \frac{1}{T}\sum_{k=1}^{N}
\int_{0}^{\infty}A(t',\xi_{k})A(t'+t,\xi_{k})\rmd t'\right\rangle +\mbox{other terms.}
\label{eq:c-intermed}
\end{equation}
In many cases  $1/f^{\beta}$ PSD is caused only by the first part. For example,
it is known that when the pulses occur randomly as a Poisson
process, the PSD of the signal depends only on the shapes of the pulses,
as given by Carson's theorem \cite{Carson1931}. Conditions
when a sequence of randomly occurring pulses leads to $1/f^{\beta}$
noise were investigated in \cite{Halford1968}. Note, that when inter-pulse
duration $\vartheta_{k}=t_{k+1}-t_{k}$ is uncorrelated with the duration of the pulse, different
pulses may be overlapping. Even in case when the other terms in equation \eref{eq:c-intermed}
are nonzero, the first part can have different dependence on time
$t$ and dominate for some time range $t_{\mathrm{min}}<t<t_{\mathrm{max}}$.
Therefore, initially we will consider only the first part of equation \eref{eq:c-intermed}.
It can be written as
\begin{equation}
C(t)=\nu\left\langle \int_{0}^{\infty}A(t',\xi)A(t'+t,\xi)\rmd t'\right\rangle \,,\label{eq:c1}
\end{equation}
where $\nu$ is the mean number of pulses per unit time. Since the pulse
duration $\tau$ is a function of the parameters $\xi$, the PDF of pulse durations is
\begin{equation}
P_{\tau}(\tau')=\int\delta(\tau(\xi)-\tau')P(\xi)\rmd\xi
\end{equation}
Introducing the autocorrelation function of the pulses with the same duration
$\tau'$, 
\begin{equation}
C(t,\tau')=\int\rmd\xi\,\delta(\tau(\xi)-\tau')\frac{P(\xi)}{P_{\tau}(\tau')}
\int_{0}^{\tau'}\rmd t'\, A(t',\xi)A(t'+t,\xi) \,,\label{eq:c-t-tau}
\end{equation}
we can write equation \eref{eq:c1} in the form
\begin{equation}
C(t)=\nu \int P_{\tau}(\tau)C(t,\tau)\rmd\tau\,. \label{eq:corr}
\end{equation}
If the PDF of pulse durations has a power-law form
\begin{equation}
P_{\tau}(\tau)\sim\tau^{\rho}\label{eq:prob-tau}
\end{equation}
and the autocorrelation function of the pulses with the same duration
has the scaling property
\begin{equation}
C(at,a\tau)=a^{2\gamma+1}C(t,\tau)\label{eq:c-tau-scaling}
\end{equation}
then it follows that the autocorrelation function $C(t)$ has the
required property \eref{eq:auto-scaling} with $\beta$ given by
equation
\begin{equation}
\beta=\rho+2\gamma+3\,\label{eq:beta}.
\end{equation}
The meaning of the parameter $\gamma$ will be revealed below,
equations~\eref{eq:h} and \eref{eq:gamma-eta}. Note, that the scaling in equation 
\eref{eq:c-tau-scaling} is the same as described by equation~(9) in
\cite{Kuntz2000}, where the pulse area $S\sim\tau^{\gamma+1}$ is used
instead of pulse duration $\tau$.

In order to avoid the divergence of the PDF of pulse durations \eref{eq:prob-tau}, equation 
\eref{eq:prob-tau} should hold only in some region of the pulse
durations $\tau$, $\tau_{\mathrm{min}}\leqslant\tau\leqslant\tau_{\mathrm{max}}$.
In this case the scaling cannot be exact. However, if the influence
of the limiting values $\tau_{\mathrm{min}}$ and $\tau_{\mathrm{max}}$
can be neglected for time $t$ in some region $t_{\mathrm{min}}\ll t\ll t_{\mathrm{max}}$,
we can expect that the scaling \eref{eq:auto-scaling} approximately holds
for this time region.

One of the ways to get the required scaling \eref{eq:c-tau-scaling}
of the autocorrelation function $C(t,\tau)$
is to consider pulses having the same shape, only stretched in height
and in time. The signal consisting of such pulses was investigated
in \cite{Halford1968}. For stretched pulses we can write
\begin{equation}
A(t,\xi)=h(\xi)y(t/\tau(\xi))\,,
\end{equation}
where $h(\xi)$ is the height of the pulse. The function $y(t_s)$ is
nonzero only when $0\leq t_s\leq 1$. 

From equation \eref{eq:c-t-tau} we obtain
\begin{equation}
C(t,\tau)=\tau\overline{h^{2}}(\tau)c(t/\tau)\,,\label{eq:c-tau-scaling-a}
\end{equation}
where
\begin{equation}
\overline{h^{2}}(\tau')=\int\delta(\tau(\xi)-\tau')h^{2}(\xi)
\frac{P(\xi)}{P_{\tau}(\tau')}\rmd\xi
\end{equation}
is the mean squared amplitude of pulses having the same duration $\tau'$
and the function 
\begin{equation}
c(t_{\mathrm{s}})=\int_{0}^{1}y(t^{\prime}_{\mathrm{s}})
y(t^{\prime}_{\mathrm{s}}+t_{\mathrm{s}})\rmd t^{\prime}_{\mathrm{s}}
\end{equation}
is the autocorrelation function of pulse shapes. If the height $h$ of
the pulse is a power-law function of the pulse duration $\tau$,
\begin{equation}
h\sim\tau^{\gamma}\,,\label{eq:h}
\end{equation}
then the scaling \eref{eq:c-tau-scaling} of the autocorrelation
function \eref{eq:c-tau-scaling-a} holds. It should be noted, that even when pulse height $h$
is proportional to $\tau^{\gamma}$, the coefficient of proportionality
is not necessarily constant. In particular, the sign of the pulses
can be random. Only coefficient of proportionality for the average
of the square of the pulse height $\overline{h^{2}}(\tau)\sim\tau^{2\gamma}$
should be constant. If pulse height is a power-law function of the
pulse duration, then the change of the magnitude of the pulse height
$h\rightarrow ah$ is caused by the change of pulse duration $\tau\rightarrow a^{1/\gamma}\tau$.
Comparing this scaling property to equation \eref{eq:prob-scaling} we
see that the power-law exponent $\gamma$ plays a similar role as $-1/\mu$, i.e., 
\begin{equation}
\gamma=-\frac{1}{\mu}=\frac{1}{2(1-\eta)}\,.\label{eq:gamma-eta}
\end{equation}
The sign minus in equation \eref{eq:gamma-eta} appears because stretching
the time, as in equation \eref{eq:prob-scaling}, is equivalent to the
shortening of the pulse duration.

Now we will investigate the influence of limiting pulse durations
$\tau_{\mathrm{min}}$ and $\tau_{\mathrm{max}}$. From the assumptions
made above, equations \eref{eq:prob-tau}, \eref{eq:beta}, and \eref{eq:h}, we have that $P_{\tau}(\tau)\overline{h^{2}}(\tau)=B\tau^{\beta-3}$
when $\tau_{\mathrm{min}}\leqslant\tau\leqslant\tau_{\mathrm{max}}$.
Here $B$ is the coefficient independent from $\tau$. For time $\tau_{\mathrm{min}}\ll t\ll\tau_{\mathrm{max}}$ we
can write the autocorrelation function, according to equations \eref{eq:corr} and \eref{eq:c-tau-scaling-a}, as
\begin{equation}
C(t)=\nu B\int_{\tau_{\mathrm{min}}}^{\tau_{\mathrm{max}}}\tau^{\beta-2}c(t/\tau)\rmd\tau
=\nu Bt^{\beta-1}\int_{1}^{\frac{\tau_{\mathrm{max}}}{t}}u^{\beta-2}c(1/u)\rmd u\,.
\label{eq:auto-approx}
\end{equation}
According to paper \cite{Halford1968}, physically reasonable pulses
are square-integrable and have everywhere finite derivative of the
autocovariance function with respect to time. Then for large $u$
we can approximate $c(1/u)\approx c(0)+c'(0)/u$ and get
\begin{equation}
C(t)\approx\cases{
\nu Bt^{\beta-1}\left[\frac{c(0)}{1-\beta}+\frac{c'(0)}{2-\beta}\right]\,,& $0<\beta<1$ ,\\
\nu B\left[c(0)\ln\tau_{\mathrm{max}}+c'(0)-c(0)\ln t\right]\,, & $\beta=1$ ,\\
\nu B\left[c(0)\frac{\tau_{\mathrm{max}}^{\beta-1}}{\beta-1}+t^{\beta-1}
\frac{c'(0)}{2-\beta}\right]\,, & $1<\beta<2$ .\\}
\end{equation}

Thus, for $0<\beta<2$ and $\tau_{\mathrm{min}}\ll t\ll\tau_{\mathrm{max}}$ the
term containing time $t$ has the scaling property the same as in equation 
\eref{eq:auto-scaling}, the limiting values of the pulse
durations $\tau_{\mathrm{min}}$ and $\tau_{\mathrm{max}}$ do not influence the scaling
of the autocorrelation function. On the other
hand, if $\beta>2$ then the influence of $\tau_{\mathrm{max}}$ becomes
significant.

\begin{table}
\caption{\label{tab:power-cases}Some situations when the power-law dependence of the pulse height $h$
on the pulse duration $\tau$ occurs. The corresponding power-law exponents $\gamma$
together with the exponents $\rho$ required to get $1/f$ PSD.}
\begin{indented}
\lineup
\item[]\begin{tabular}{@{}llll}
\br
Signal&\m$\gamma$&\m$\rho$&Meaning of $\rho$\cr
\mr
Constant area pulses&$-1$&\m$0$&Uniform distribution of pulse durations\cr
Constant energy pulses&$-1/2$&$-1$&\cr
Pulses of constant height&\m$0$&$-2$&Uniform distribution of inverse durations\cr
Geometrically similar pulses&\m$1$&$-4$&\cr
\br
\end{tabular}
\end{indented}
\end{table}

As has been pointed out in \cite{Ruseckas2003}, the condition for
$1/f$ spectrum, $\rho+2\gamma+2=0$, can be easily satisfied. The power-law
dependence of the pulse height on the pulse duration can occur naturally. 
Various cases are listed in table~\ref{tab:power-cases}. The value $\gamma=0$ corresponds to
the pulses of constant height; $\gamma=-1$ corresponds to constant area pulses.
Geometrically similar pulses have $\gamma=1$. If the energy is proportional to square of
the signal, the constant energy pulses correspond to $\gamma=-1/2$. Since
we have $1/f$ spectrum when $\rho=-2(\gamma+1)$, this spectrum occurs
for constant area pulses ($\gamma=-1$) and uniform distribution of
pulse durations ($\rho=0$) in a wide interval. For constant height
pulses ($\gamma=0$) we have $1/f$ spectrum when the distribution
of inverse durations $\tau^{-1}$ is uniform, that is when $P_{\tau}(\tau)\propto\tau^{-2}$.

Signal consisting of overlapping constant height pulses has PDF of
Poisson distribution. On the other hand, pulses with $\gamma\neq0$
can lead to power-law tails in the PDF of the signal. Let us consider
rectangular pulses with the only random parameter being the pulse
duration $\tau$. Large signal intensities are due to pulses of large
height, for which one can neglect the overlap between pulses. Each
pulse of the height $h(\tau)$ occurs with the probability $P_{\tau}(\tau)$
and lasts for time $\tau$. Thus the PDF of the signal intensity $I=h$
is (see for analogy \cite{Kaulakys2005}) 
\begin{equation}
P_{I}(I)=\frac{\tau}{\langle\tau\rangle}P_{\tau}(\tau)
\left.\frac{\rmd\tau}{\rmd h}\right|_{h=I}\,.\label{eq:pdf-1}
\end{equation}
If the PDF of pulse durations has the power-law $P_{\tau}(\tau)\propto\tau^{\rho}$ form
and the height of the pulse depends on the pulse duration as $h\propto\tau^{\gamma}$,
then from equation \eref{eq:pdf-1} we obtain $P_{I}(I)\propto I^{-\lambda}$,
where
\begin{equation}
\lambda=1-\frac{2+\rho}{\gamma}\,.\label{eq:lambda-pulses}
\end{equation}

For the pure $1/f$ noise $2+\rho=-2\gamma$ and we get the exponent
$\lambda=3$. For the case of $f^{-\beta}$ spectrum we have the following
relation between the exponent $\beta$ of the spectrum and exponent
$\lambda$ of the signal PDF:
\begin{equation}
\beta=1+\gamma(3-\lambda)\,.\label{eq:beta-pulses}
\end{equation}
Taking into account equation \eref{eq:gamma-eta} we see that relation \eref{eq:beta-pulses}
is the same as by equation \eref{eq:beta-sde} describing the power-law spectrum
of the signal generated by the nonlinear SDE 
\eref{eq:sde}. Note that PDF of the signal intensity has the
same power-law exponent $\lambda=3$ also when $1/f$ noise is generated
by the nonlinear SDE \eref{eq:sde}.

\begin{figure}
\includegraphics[width=0.33\textwidth]{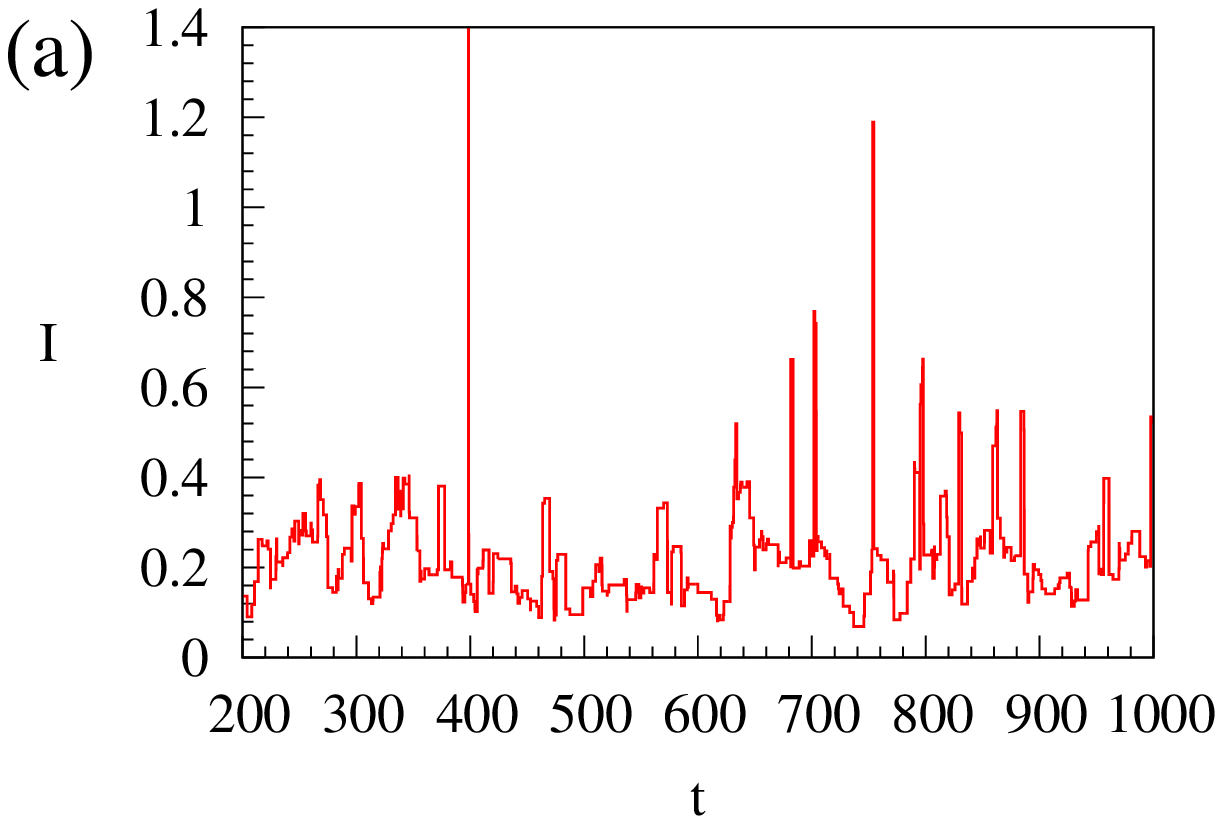}\includegraphics[width=0.33\textwidth]{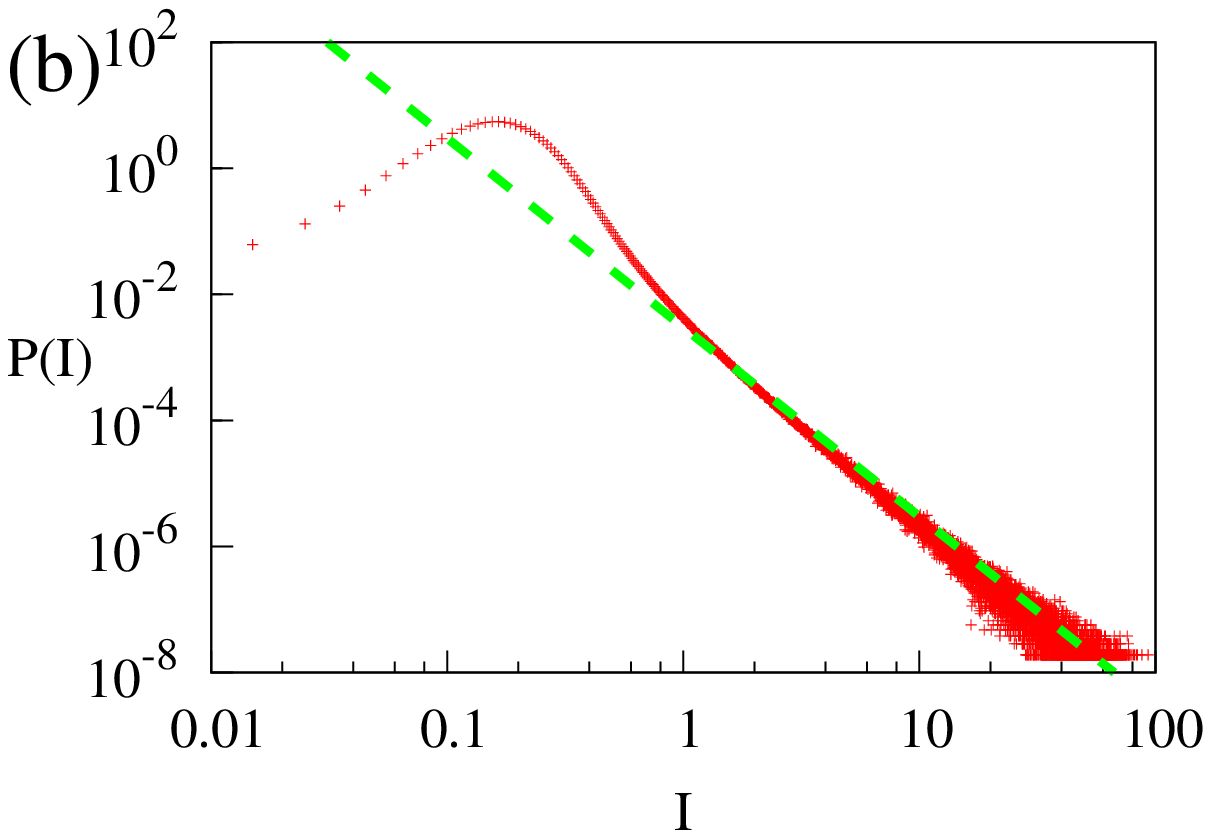}\includegraphics[width=0.33\textwidth]{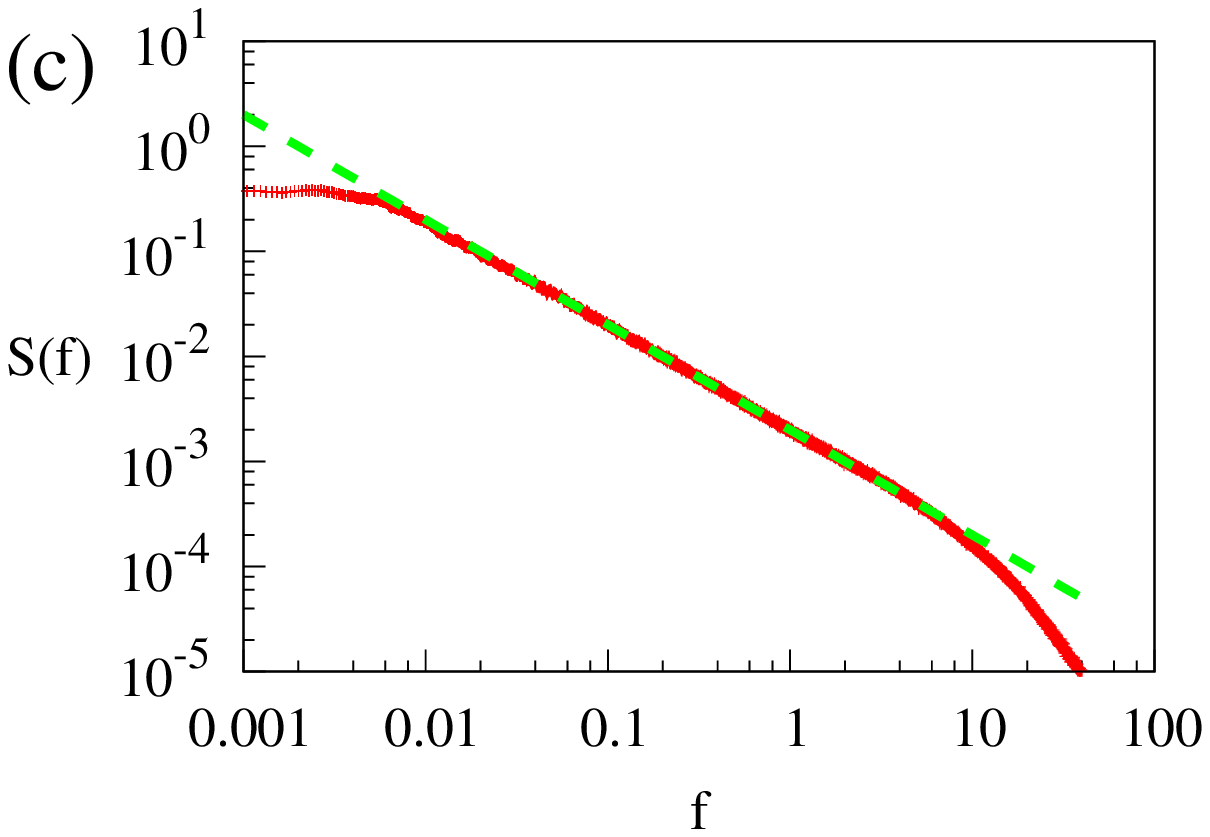}
\caption{(Color online) (a) Typical signal consisting from the constant area
rectangular pulses with the uniformly distributed durations, $\rho=0$. (b) The PDF
of the signal intensity. The dashed (green) line shows the power-law
with the exponent $-3$. (c) The PSD of such a signal. The dashed
(green) line corresponds to the spectrum obtained using Wiener-Khintchine
theorem from the autocorrelation function. The time intervals between
the pulses are distributed according to Poisson process with the average
$\bar{\vartheta}=5$ . The parameters used are $\tau_{\mathrm{min}}=0.01$ and 
$\tau_{\mathrm{max}}=100$.}
\label{fig:fixed-area}
\end{figure}

Typical signal for rectangular constant area pulses ($\gamma=-1$) is shown in
figure~\ref{fig:fixed-area}a, the PDF of the signal is shown in
figure~\ref{fig:fixed-area}b and the PSD in figure~\ref{fig:fixed-area}c. We see
a good agreement of numerically obtained PSD with the analytical estimation. In
figure~\ref{fig:fixed-area}b we can see that the PDF has a power-law tail for
large signal intensities. Note that due to the overlapping of the pulses the
PDF of the signal at smaller intensities is not a power-law and acquires a
power-law tail only for larger intensities, where the overlap can be neglected.
This is in contrast with the SDE \eref{eq:sde}, where the steady state PDF of
the signal can have a power-law form for all values of the signal sufficiently
far from limiting values $x_{\mathrm{min}}$ and $x_{\mathrm{max}}$.

\section{Connection between the nonlinear SDE and stochastic pulse sequences}

\label{sec:sde-pulses}As it was shown in Secs.~\ref{sec:SDE} and
\ref{sec:pulses}, the signals having $1/f^{\beta}$ PSD and generated
by the nonlinear SDEs have similar scaling like the signals consisting from random pulses. 
In this Section we approximate nonlinear SDE by a sequence
of pulses and show that this approximation gives the same PDF of signal
intensity and a power-law region in the PSD with the same exponent.
However, other details of the approximated signal can be different: for example,
the signal consisting from pulses does not exhibit the intermittent bursts characteristic
for solution of SDE.

Let us consider SDE \eref{eq:sde} together with reflective boundaries
at $x=x_{\mathrm{min}}$ and $x=x_{\mathrm{max}}$. Using Euler-Maruyama
approximation with time step $\Delta t=s$ and replacing the stochastic
differential equation with the difference equation we have 
\begin{equation}
x_{k+1}=x_{k}+\sigma^{2}\left(\eta-\frac{\lambda}{2}\right)x_{k}^{2\eta-1}s+\sigma x_{k}^{\eta}\sqrt{s}\varepsilon_{k}\,.
\end{equation}
Here $\varepsilon_{k}$ is a Gaussian random variable with zero mean
and unit variance. Variable time step \cite{Kaulakys2006}
\begin{equation}
s_{k}=\frac{\kappa^{2}}{\sigma^{2}}x_{k}^{2(1-\eta)}\label{eq:hk}
\end{equation}
results in the equation
\begin{equation}
x_{k+1}=x_{k}\left(1+\frac{1}{2}\kappa^{2}(2\eta-\lambda)+\kappa\varepsilon_{k}\right)\,,\label{eq:discrete}
\end{equation}
where $\kappa\ll1$ is a small parameter. The approximation of SDE
becomes better with decreasing $\kappa$. It should be noted that equation 
\eref{eq:discrete} has an universal form: $x_{k}$ enters only
in the first power independent on the exponent $\eta$. We can estimate
the steady state PDF of $x_{k}$ in $k$-space by considering equation \eref{eq:discrete}
as an Euler-Maruyama approximation of the SDE in $k$-space, 
\begin{equation}
\rmd x=\kappa^{2}\left(\eta-\frac{\lambda}{2}\right)x\rmd k+\kappa x \rmd W_{k}\,.
\label{eq:xk}
\end{equation}

Taking into account reflective boundaries at $x_{\mathrm{min}}$ and
$x_{\mathrm{max}}$ we get from the Fokker-Planck equation in $k$-space corresponding
to equation \eref{eq:xk} that the steady state PDF has the power-law
form, $P(x_{k})\propto x_{k}^{2(\eta-1)-\lambda}$. From this steady state
PDF and equation \eref{eq:hk} it follows that PDF of $s_{k}$ has
a power-law form $P_k(s_{k})\propto s_{k}^{\rho}$ with
\begin{equation}
\rho=\frac{\lambda-1}{2(\eta-1)}-2\,.\label{eq:rho-x}
\end{equation}
The same expression for the power-law exponent $\rho$ can be obtained
from equations \eref{eq:gamma-eta} and \eref{eq:lambda-pulses}.

From equation \eref{eq:discrete} we obtain $x_{k+n}$:
\begin{equation}
x_{k+n}=x_{k}\prod_{i=1}^{n}\left(1+\frac{1}{2}\kappa^{2}(2\eta-\lambda)
+\kappa\varepsilon_{k+i-1}\right)\approx x_{k}(1+\kappa\sqrt{n}\varepsilon+\cdots)\,,
\end{equation}
where $\varepsilon$ is a Gaussian random variable with zero mean
and unit variance. Here we used the fact that the sum of $n$ Gaussian
variables $\varepsilon_{k+i-1}$ is a Gaussian variable with the dispersion
equal to $n$. We can conclude that $x_{k+n}$ does not differ significantly
from $x_{k}$ as long as $\kappa\sqrt{n}\ll1$. The maximal value
of $n$ when $x_{k+n}$ is approximately equal to $x_{k}$ is $n_{\mathrm{max}}\sim1/\kappa^{2}$.
The duration in which the stochastic variable $x$ does not change
significantly is
\begin{equation}
\tau(x_{k})=n_{\mathrm{max}}s_{k}=\frac{1}{\sigma^{2}}x_{k}^{2(1-\eta)}=\frac{1}{\sigma^{2}}x_{k}^{1/\gamma}\,,\label{eq:tau-x}
\end{equation}
where $\gamma$ is given by \eref{eq:gamma-eta}.
The duration $\tau$, being proportional to $s_{k}$, has the power-law
PDF with the same exponent $\rho$ as the PDF of $s_{k}$:
\begin{equation}
P_{\tau}(\tau)=\cases{
C\tau^{\rho}\,, & $\tau_{\mathrm{min}}\leq\tau\leq\tau_{\mathrm{max}}$ ,\\
0\,, & otherwise .\\}
\label{eq:pdf-tau}
\end{equation}
Here $C$ is normalization coefficient and
\begin{equation}
\tau_{\mathrm{min}}=\frac{1}{\sigma^{2}x_{\mathrm{max}}^{2(\eta-1)}}\,,\qquad\tau_{\mathrm{max}}=\frac{1}{\sigma^{2}x_{\mathrm{min}}^{2(\eta-1)}}\,. 
\end{equation}
From equation \eref{eq:tau-x} the value of the stochastic variable $x_{k}$ is connected with the
duration $\tau$ by the relation 
\begin{equation}
x_{k}=\sigma^{2\gamma}\tau^{\gamma}\label{eq:x-tau}\,.
\end{equation}

Therefore, we can approximate the signal generated by SDE \eref{eq:sde} by rectangular
pulses of random duration $\tau$ having the PDF of durations \eref{eq:pdf-tau}
and pulse height $h\equiv x_k$ related to the pulse duration $\tau$ by equation \eref{eq:x-tau}.
The pulses are not overlapping and immediately follow each other.
Although the durations of adjacent pulses obtained from the signal
generated by SDE \eref{eq:sde} are correlated, for simplicity we will 
neglect this correlation. The PDF of the signal $x$ constructed as
such a pulse sequence has power-law form. Using equations \eref{eq:lambda-pulses}
and \eref{eq:rho-x} we get that the power-law exponent in the PDF
$P(x)$ is equal to $-\lambda$, with $\lambda$ appeared in SDE \eref{eq:sde}. 

When pulses occur not randomly but follow each other, the other terms
in equation \eref{eq:c-intermed} are nonzero. However, one can check
that for some range of time $t$ the first part in equation \eref{eq:c-intermed}
dominates. Thus the PSD of this pulse sequence has a power-law
part with the exponent given by equation \eref{eq:beta-pulses}. Using
the value of the exponent $\gamma$ from equation \eref{eq:gamma-eta}
we get the power-law exponent \eref{eq:beta-sde} in the PSD. The
frequency range $\tau_{\mathrm{max}}^{-1}\ll f\ll\tau_{\mathrm{min}}^{-1}$
where PSD of the signal consisting of pulses has power-law behavior
coincides with inequalities \eref{eq:freq-range}. Thus, the proposed approximation
of SDE by the sequence of pulses correctly reproduces power-law parts
of the PDF and the PSD of the generated signal.

\begin{figure}
\includegraphics[width=0.33\textwidth]{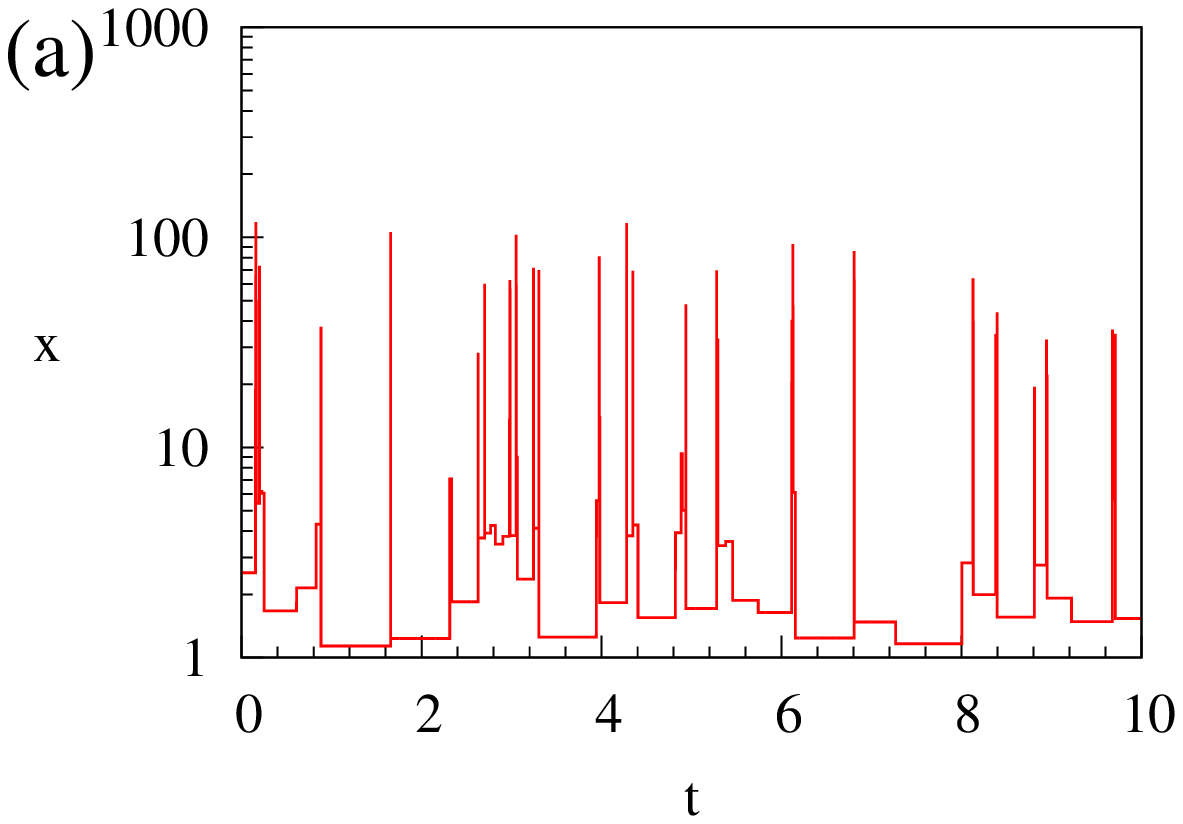}\includegraphics[width=0.33\textwidth]{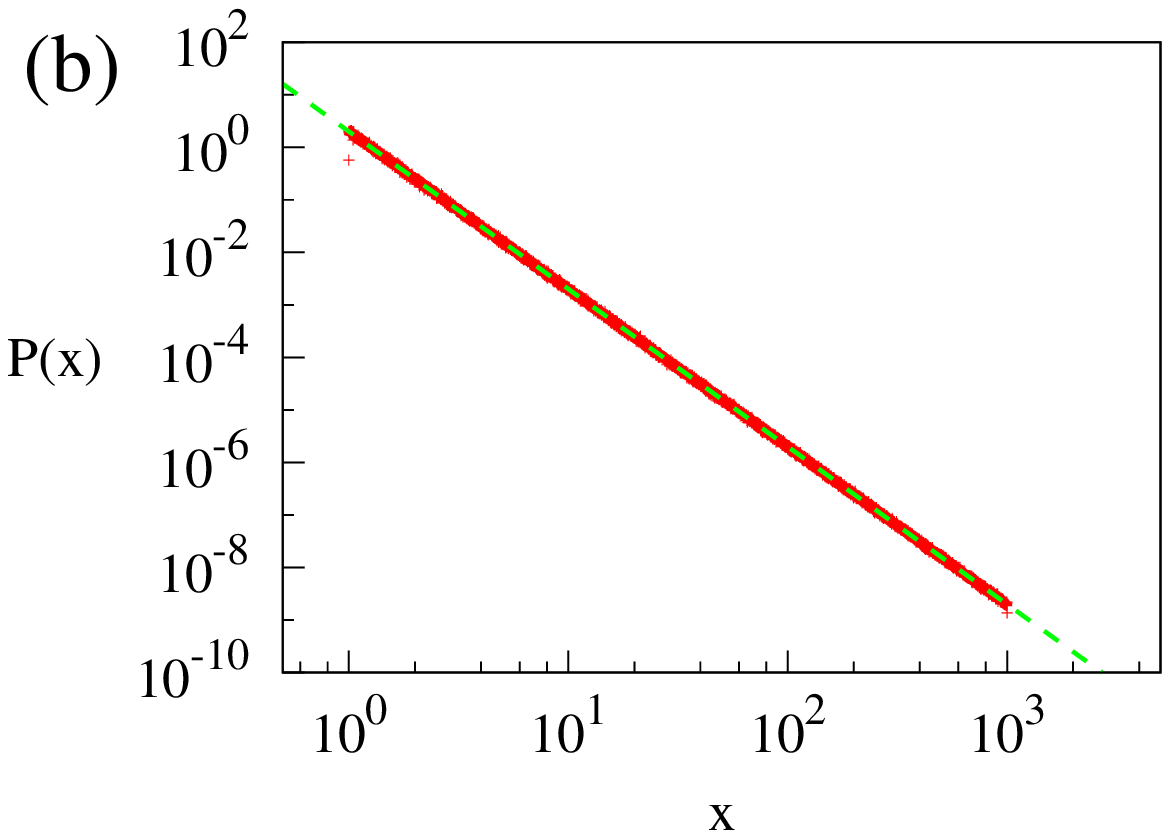}\includegraphics[width=0.33\textwidth]{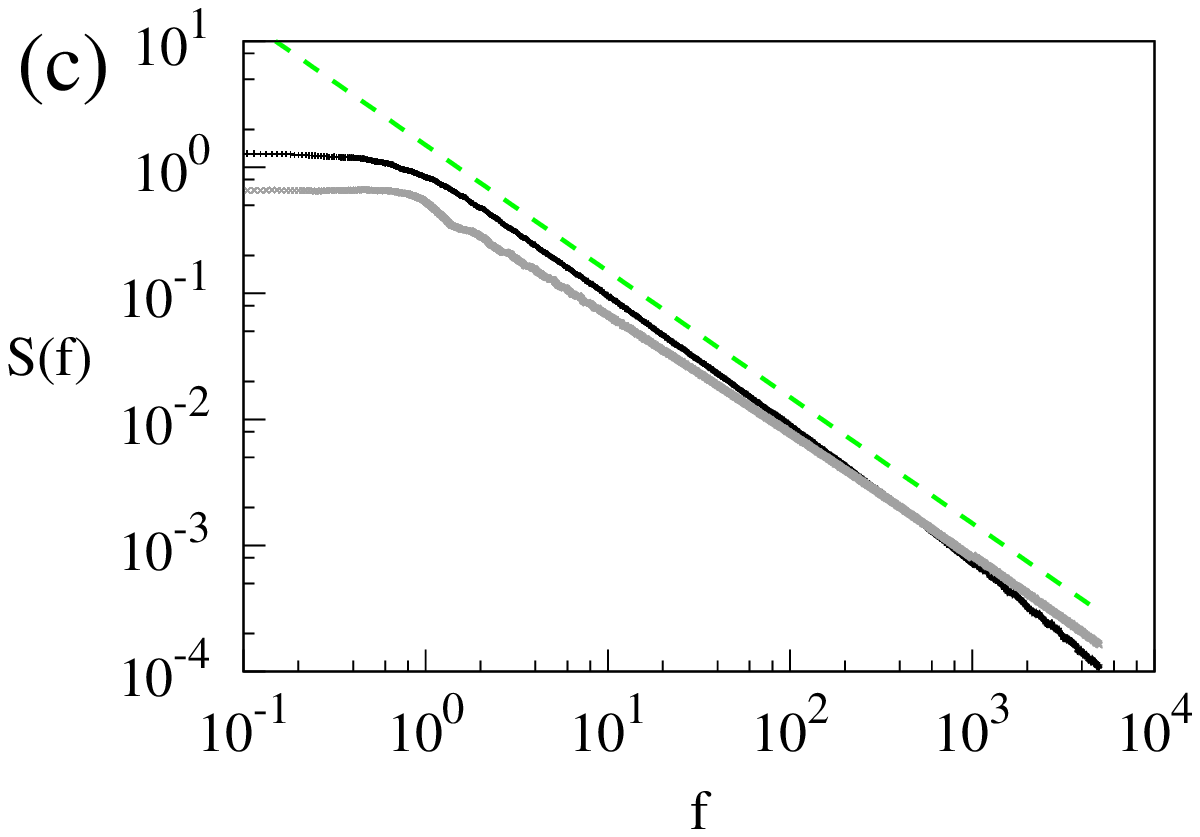}
\caption{(Color online) (a) Typical signal consisting of the equal energy pulses that immediately follow each other. The PDF of pulse durations $\tau$ is given by equation 
\eref{eq:pdf-tau} with $\rho=-1$, $\tau_{\mathrm{min}}=10^{-6}$, and 
$\tau_{\mathrm{max}}=1$ and pulse heights are proportional to $\tau^{-1/2}$.
(b) The PDF of the signal intensity. The dashed (green) line shows
the power-law with the exponent $-3$. (c) Comparison of the PSD calculated
using the signal generated by SDE \eref{eq:sde} with the same parameters
as in figure~\ref{fig:sde} (black line) and using the signal consisting
of pulses (gray line). The dashed (green) line shows the slope $1/f$.}
\label{fig:cmp}
\end{figure}

To illustrate the approximation of nonlinear SDE by a sequence of
pulses, in figure~\ref{fig:cmp} we compare the PDF and the PSD of
the signal consisting from pulses to the PDF and the PSD of the signal
shown in figure~\ref{fig:sde}. The SDE has parameters $\eta=2$, $\lambda=3$,
$x_{\mathrm{min}}=1$, $x_{\mathrm{max}}=1000$, therefore, duration
of the pulses has power-law PDF \eref{eq:pdf-tau} with $\rho=-1$
(according to equation \eref{eq:rho-x}) and $\tau_{\mathrm{min}}=10^{-6}$,
$\tau_{\mathrm{max}}=1$. The height of each pulse is proportional
to the duration of the pulse to the power of $\gamma=-1/2$, obtained
from equation \eref{eq:gamma-eta}. This value of $\gamma$ means that
each pulse has the same energy. The signal consisting of such pulses
is shown in figure~\ref{fig:cmp}a. As one can see, this signal looks
rather different from the one shown in figure~\ref{fig:sde}a. Large values
of the signal in figure~\ref{fig:cmp}a do not come in the intermittent bursts as in
figure~\ref{fig:sde}a. This difference
is caused by the assumption that the durations of different pulses
are uncorrelated. The PDF of the signal, shown in \ref{fig:cmp}b,
is the same as in figure~\ref{fig:sde}b. Comparison of the PSDs is
shown in figure~\ref{fig:cmp}c. There is qualitative agreement between
the PSD of the signal generated by the nonlinear SDE and the PSD of the
signal consisting of pulses. The PSD of the signal consisting of pulses
has a power-law part in a different range
of frequencies, from $f\approx10^{1}$ up to $f\approx4\times10^{4}$.
This difference from the expected range
$\tau_{\mathrm{max}}^{-1} < f < \tau_{\mathrm{min}}^{-1}$
is caused by neglected other terms in equation \eref{eq:c-intermed}.

\section{Conclusions}

\label{sec:concl}In summary, we have demonstrated the connection
between the nonlinear SDEs generating signals with $1/f^{\beta}$ noise
and signals consisting of random pulses with the power-law distribution
of pulse durations. The exponent $\rho$ of the power-law PDF of pulse
durations and the exponent $\gamma$ characterizing the dependence
of the pulse height on the pulse duration are related to the parameters
$\eta$ and $\lambda$ of the SDE \eref{eq:sde} by means of the equations 
\eref{eq:gamma-eta} and \eref{eq:rho-x}. The signal generated
by SDE and corresponding signal consisting of rectangular pulses yield
the same distribution of signal intensity and the same power-law exponent
in the PSD \eref{eq:beta-sde}. The appearance of $1/f^{\beta}$
spectrum and relationship between parameters can be obtained just
by considering the scaling properties of the signals. The revealed
connection between different models of $1/f$ noise provides further
insights into the origin and relationship between different models of $1/f$ noise. 

\section*{References}

\providecommand{\newblock}{}

\end{document}